\documentclass{article}

\usepackage{graphicx,amsmath,authblk}
\usepackage{bbold}
\graphicspath{{figures/}}

\usepackage{geometry}
\begin{document}
\title{Single pixel structured imaging through fog}
\author[1]{Mark Bashkansky\footnote{bashkansky@nrl.navy.mil}}
\author[1]{Samuel D. Park}
\affil[1]{U.S. Naval Research Laboratory, 4555 Overlook Ave., SW, Washington, DC 20375}

\author[2]{John Reintjes}
\affil[2]{Jacobs Technology, Inc., 2551 Dulles View Dr., Suite 700, Herndon, VA  20171}

\maketitle

\begin{abstract}
We describe the application of structured imaging with a single pixel camera to imaging through fog.  We demonstrate the use of a high-pass filter on the detected bucket signals to suppress the effects of temporal variations of fog density and enable an effective reconstruction of the image. A quantitative analysis and comparison of several high-pass filters are demonstrated for the application. Both computational ghost imaging and compressive sensing techniques were used for image reconstruction and compressive sensing was observed to give a higher reconstructed image quality.
\end{abstract}

\section{Introduction}
Imaging with structured light beams using single pixel cameras has attracted considerable attention in recent years. Both ghost imaging and compressive imaging approaches have been implemented. Applications have included 3-D imaging \cite{li2012gated,hardy_computational_2013,sun_single-pixel_2016}, single photon imaging \cite{zhu_photon-limited_2020}, communication \cite{cox_structured_2021}, imaging through turbulent \cite{hardy_reflective_2011,erkmen_computational_2012,shirai_imaging_2012,yao_effect_2013} and stationary scattering media \cite{liutkus_imaging_2015,duran_compressive_2015,rajaei_intensity-only_2016}. The use of a high-pass digital filter was described to reduce the effects of background light disturbances in ghost imaging \cite{zhang_digital_2021}.

In this paper we consider the application of structured light beams to imaging through fog. The major issue with imaging through fog is attenuation due to scattering, which presents problems for all imaging systems. In this regard, single-pixel imaging systems have a potential advantage because they focus all of the available photons onto a single detector, increasing the signal to noise ratio (SNR). However, there are additional problems in imaging through fog even without turbulence, which is usually responsible for image degradation over long distances. One prominent issue is the temporal variation of the detected signal due to unavoidable changes in the fog density with time. These changes can be confused with temporal variations in the detected signal associated with changes in the target reflectivity as the structure on the illumination beam is varied, thus reducing the fidelity of the reconstructed image. Another issue is the presence of light scattered by fog itself. This is addressed by using short pulses in combination with time gating of fast detector. We describe here a mitigation technique for the temporal variations of the signal due to variation in the fog density. It is based on the observation that the time scale of the intensity variations of gated detector due to changes in fog density is much slower than the time scale for the variations due to changes in the structured beam illumination. As a result, it becomes possible to suppress the effects of fog density variations by using fast pulses, a fast detector and a high-pass filter in the detection system. Using this technique, we demonstrate that the image of the target can be successfully reconstructed in the presence of fog using both computational ghost imaging and compressive sensing, which has the potential for shorter acquisition times. 

Our technique uses high-pass filters similar to the technique described in ref. \cite{zhang_digital_2021}. However, in their application, the source of the background fluctuations was external to the illumination-imaging system and can be mitigated by other means, such as optical narrowband filtering or use of a reference detector. In contrast, the temporal fluctuations in our application are introduced directly onto the returned signal, making it impossible to compensate for them with a simple reference detector. 

\section{Experimental Setup}
\begin{figure}[h!]
\centering\includegraphics[scale=0.75]{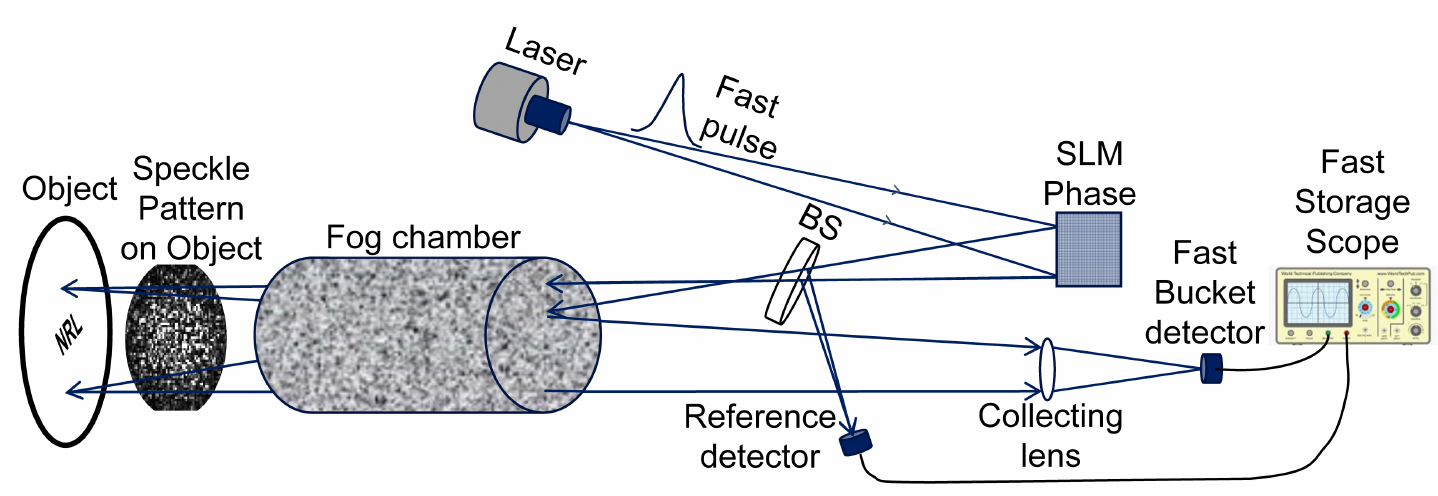}
\caption[]{ Simplified experimental diagram showing the pulsed laser source, the spatial light modulator (SLM), the two pass fog chamber, the target, the collecting lens and fast bucket detector.  Beam splitter (BS) is used to monitor the total power in the illumination beam before the fog chamber to allow for correction due to fluctuation in laser pulse intensity. The signal is recorded on a fast storage scope. }
\label{fig:exptapp}
\end{figure}

The experimental set up is shown in Fig. \ref{fig:exptapp}. The illumination signal is formed by a pulsed laser followed by a spatial light modulator (SLM) that imposes a random phase pattern on the illumination beam.  The illumination beam propagates through a fog chamber and develops a speckled intensity distribution in its far field at the target. The beam that is reflected from the target propagates back through the fog chamber and is focused by a lens onto a fast bucket detector and then recorded on a fast storage scope.

The fog chamber is custom-built using Lexan\textregistered{} polycarbonate sheets (dimensions: 8x10x48 inches). Anti-reflective coated optical windows are used for the beam path. The fog is generated using a commercial fog machine (Nutramist cyclone ultrasonic fogger) and a ball valve is used to regulate the rate of introduction of the fog. The fog enters the chamber through holes equidistantly spaced in a polyvinyl chloride (PVC) pipe extending the length of the fog chamber; This technique results in an approximately uniform fog distribution throughout the entire chamber.

The laser is a Bright Solutions SRL wedge XF 532 that produces 10 $\mu$J 532-nm sub-nanosecond pulses at a repetition rate of 10 kHz. The Meadowlark SLM produces random speckle patterns in the far field that are updated at a 40 Hz rate.  A beam splitter reflects a portion of the illumination light onto a reference detector to normalize pulse-to-pulse fluctuations in the laser energy.  We average 100 laser shots for each speckle pattern to reduce laser fluctuations further.

\begin{figure}[h!]
\centering\includegraphics[scale=1.0]{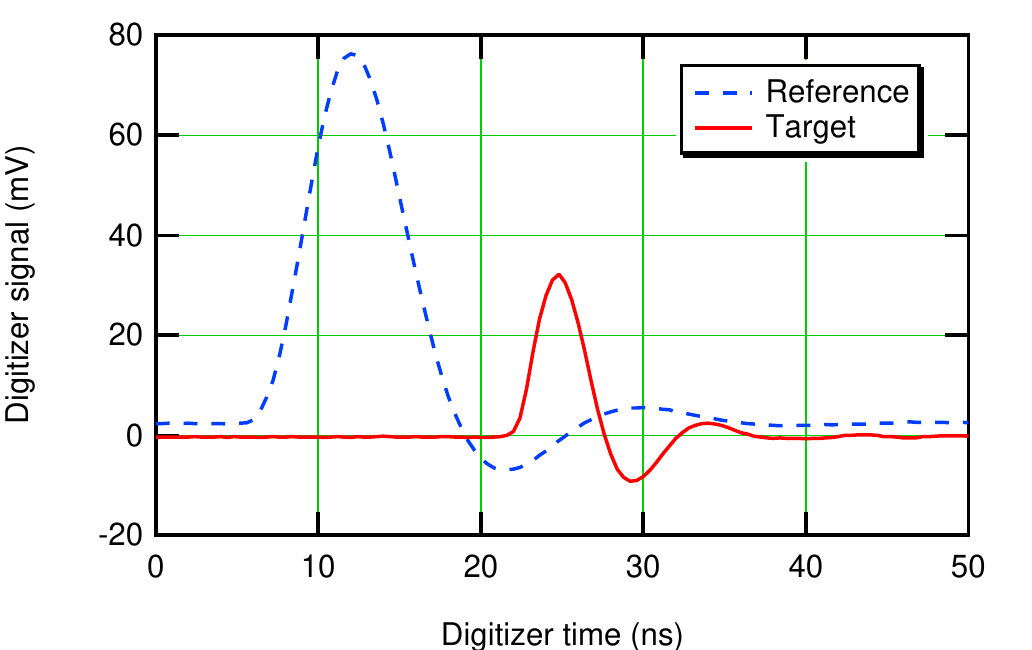}
\caption[]{ Detected bucket signal as function of time without fog. }
\label{fig:buckettrace}
\end{figure}

The diffuse target, a cutout of the letters NRL on white paper, is placed just outside the 4-foot-long fog chamber. The double pass attenuation of light in the fog was exp(-6), measured using a separate 532 nm continuous wave (CW) laser propagating through the fog chamber (not shown in figure). Higher attenuations could be used in principle, but under those conditions the fog scatter near the target exceeded the dynamic range of our digital oscilloscope. We collected a number of traces from the bucket detector (Thorlabs\textregistered{} DET10A2) using a PicoScope\textregistered{} 6407 Digitizer. Gated values of the trace, corresponding to the distance to the object, were used together with the stored speckle patterns for reconstruction of the image (NRL) with computational ghost imaging (CGI) and compressive sensing imaging (CSI). An example of a typical time signal obtained without fog (red solid curve) along with a reference signal (blue dashes) for normalization and range is shown in Fig. \ref{fig:buckettrace}. Collecting the full trace enables the ability to differentiate and reconstruct multiple targets located at different distances.

\section{Image Reconstruction}
The image is reconstructed using an the SLM and a single-pixel (bucket) detector. The SLM is used to project unique 2D intensity patterns $I_i(x,y)$ onto the object and the reflection intensity is measured using a the single-pixel detector,

\begin{equation}
    b_i = \iint I_i(x,y)O(x,y) \space dxdy,
\end{equation}
where $b_i$ is the peak of the integrated time signal and $O(x,y)$ is the reflection function of the object/target. Provided that the bucket detector is collecting the scattered light from the object illuminated by the 2D intensity pattern, $b_i$ is effectively a time-gated weighting factor for each unique intensity pattern in the image reconstruction algorithm.

To reconstruct the object’s reflection function using CGI, the 2-D intensity patterns are weighted with the bucket detector measurements \cite{katz_compressive_2009},

\begin{equation}
    O(x,y) = \langle (b_i - \langle b \rangle)I_i(x,y)\rangle,
\end{equation}
where $\langle \cdot \rangle = \frac{1}{M}\sum \cdot$ denotes an ensemble average over all the measurements, M. An alternative algorithm that subtracts the mean of the bucket detector signal has also been employed for an improvement in the image reconstruction \cite{ferri_differential_2010}

\begin{equation}
    O_{CGI}(x,y) = \langle b_i - \langle b \rangle)(I_i(x,y) - \langle I(x,y) \rangle ) \rangle.
\label{eqn:cgi}
\end{equation}

\begin{figure}[h!]
\centering\includegraphics[scale=0.6]{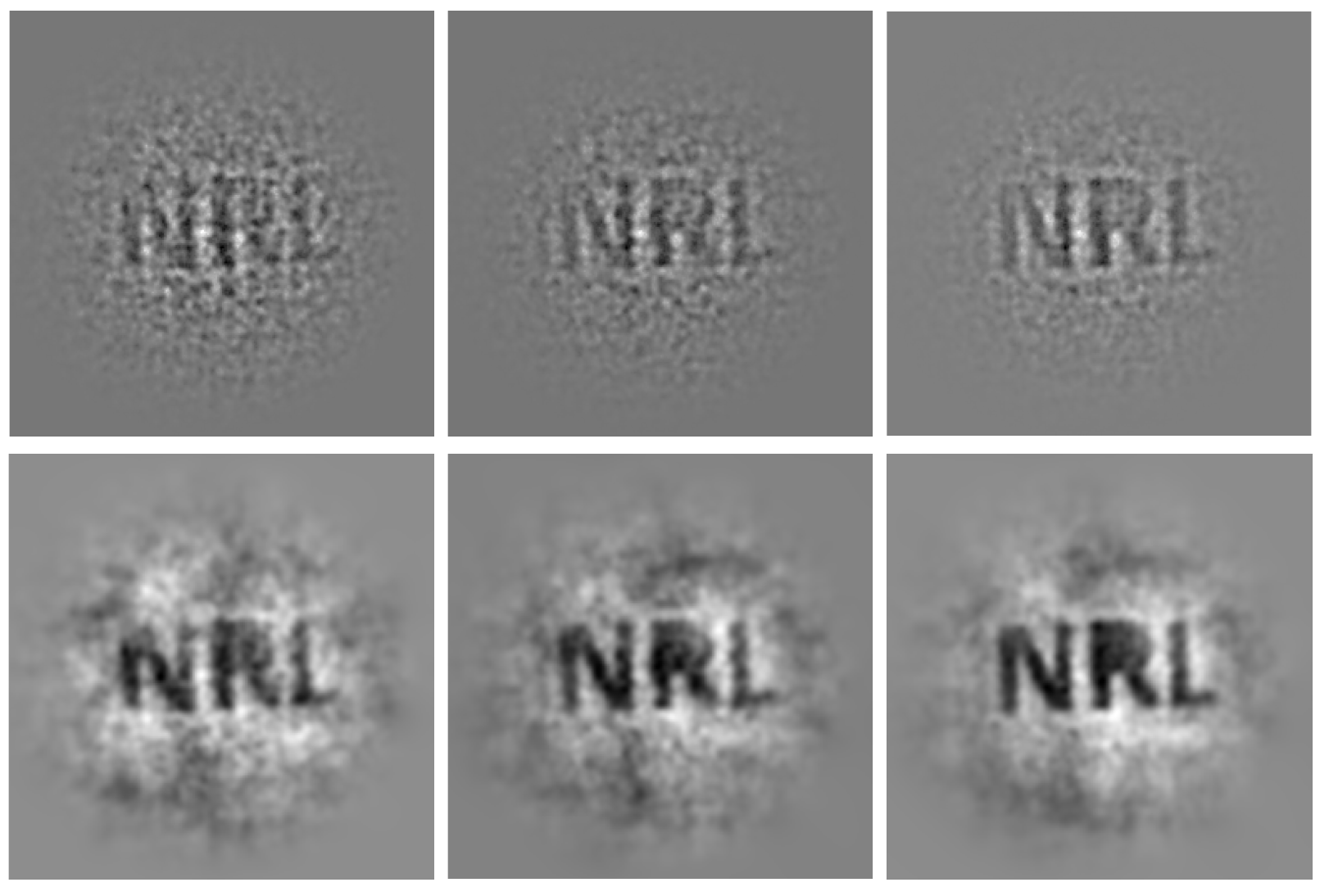}
\caption[]{ Comparison between CGI (top row) and CSI (bottom row) reconstructed images without fog of the target with varying number of speckle measurements. The first, second, and third columns show observations of 1000, 2000, and 4000 speckle realizations. Each image has a resolution of 256 x 256 pixels.}
\label{fig:cgi_csi_compare}
\end{figure}

For CSI we employ TVAL3\cite{li_efficient_2013}, a total variation (TV) minimization solver, due to its speed and robustness in the presence of noise. The compressive sensing image reconstruction is performed by solving for the object image using the model,
\begin{equation}
\min_{O_{CSI}} \sum_i \lvert\lvert D_i O_{CSI} \rvert\rvert + \frac{\mu}{2} \lvert\lvert AO_{CSI} - b\rvert\rvert^2_2
\end{equation}
where $D_iO_{CSI}$ is the discrete gradient, or total variation, of the reconstructed image $O_{CSI}$ at pixel $i$, $A$ is the measurement matrix (speckle realizations), and $b$ is the compressed signal, or bucket values. The parameter $\mu$ is the penalty parameter that is adjusted to compensate for the noise in the bucket values and the sparsity level of the reconstructed image $O_{CSI}$. The image reconstruction comparison between CGI and CSI is shown in Figure \ref{fig:cgi_csi_compare} with varying number of speckle realizations. Both techniques are able to resolve the NRL target, but the CSI reconstruction using TVAL3 shows an improvement in contrast while preserving the edges of the image. It is worth noting that the image can be adequately reconstructed with CSI using fewer speckle realizations, providing the potential for faster data acquisition at the expense of computation time.

\section{Measurements with fog}
\begin{figure}[h!]
\centering\includegraphics[scale=1.25]{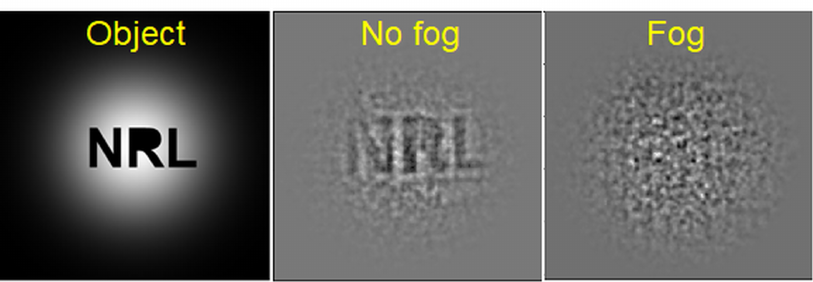}
\caption[]{ 2-D camera image of the object, CGI reconstruction without fog and CGI reconstruction with fog using 4000 speckle patterns. }
\label{fig:cgi_fog}
\end{figure}

Figure \ref{fig:cgi_fog} shows the CGI image reconstruction using Eqn. \ref{eqn:cgi} with and without fog using 4000 speckle patterns. The middle panel shows the image reconstruction without fog. Although the target is not completely resolved, the NRL letters are distinguishable. However, the reconstruction with fog using the same approach completely fails as can be seen in the right panel.

\begin{figure}[h!]
\centering\includegraphics[scale=1]{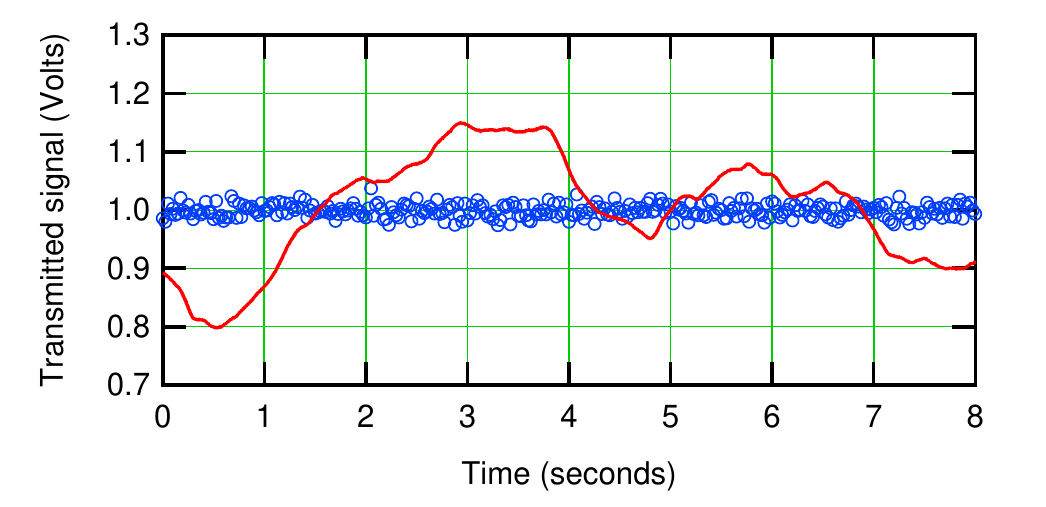}
\caption[]{Time varying transmission through fog chamber. Red solid curve shows transmission of CW light. Blue circles show pulsed bucket detector measurements used in structured light reconstruction without fog. }
\label{fig:fogtrace}
\end{figure}

A potential source of the failed reconstruction is the additional temporal fluctuations in the bucket signal introduced by the fog that are not associated with the changing SLM patterns. The time varying signal for CW laser light transmitted through the fog chamber is shown as red curve in Fig. \ref{fig:fogtrace} along with the bucket detector measurements used without fog, shown as the blue circles.

It can be seen that the transmitted laser intensity displays peak-to-peak variations of up to 30\%.  This does not present significant issues for single-shot or multi-shot focal plane array detectors. However, structured light imaging relies on small changes in the detected light for multiple single detector exposures.  These desired fluctuations are comparatively small, on the order of 1\%. The ability to reconstruct an image using structured light will be severely degraded by any noise added to the bucket detector measurements. It can be expected that the additional noise fluctuations due to fog shown in Fig. \ref{fig:fogtrace} will render reconstruction impossible with the usual techniques.

\begin{figure}[h!]
\centering\includegraphics[scale=1]{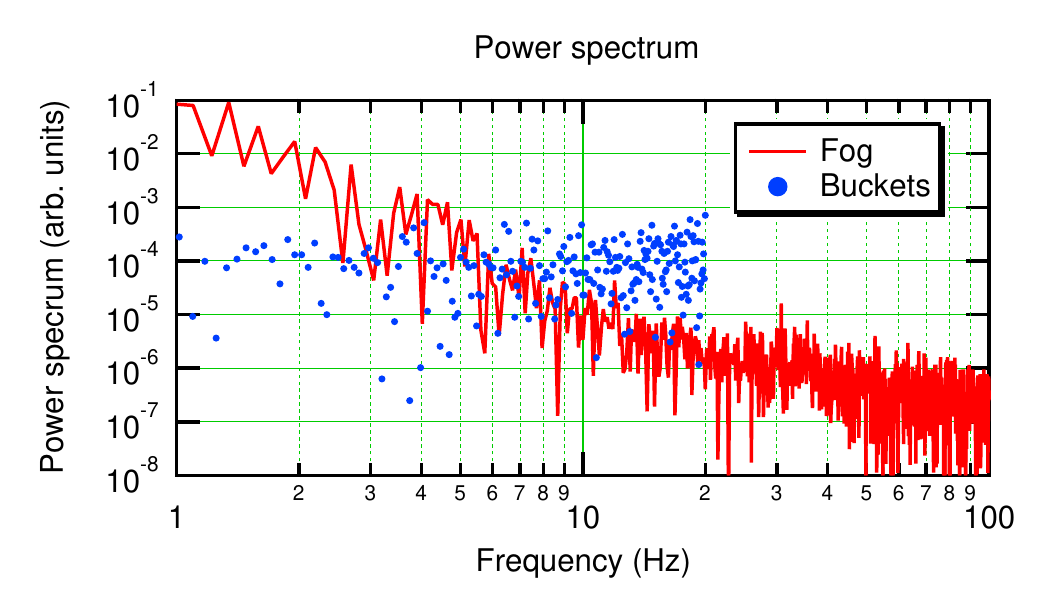}
\caption[]{Red solid curve shows power spectrum of light transmitted through the fog chamber. Blue dots show power spectrum of the bucket detector measurements without fog. }
\label{fig:fog_power}
\end{figure}

However, it is also evident from Fig. \ref{fig:fogtrace} that the fog-induced power fluctuations occur on a significantly slower time scale than the variations associated with the changing speckle patterns. This is confirmed by examining the power spectrum of the signals as shown in Fig. \ref{fig:fog_power}.

Our bucket detector measurements were taken at 40 Hz, with a Nyquist frequency of 20 Hz. As can be seen in Fig. \ref{fig:fog_power} most of the fog noise appears below $\sim$6 Hz.  We can then expect that high-pass filtering of the bucket signal through the fog will remove most of the noise and only some of the needed bucket fluctuations, enabling reconstruction of the structured light even in the presence of fog.  A successful demonstration of this technique is given in the next section.  A similar situation was described in Ref. \cite{zhang_digital_2021}, where slow fluctuations in the background intensity were shown to degrade the image reconstruction.  As mentioned earlier a high-pass filter was used in Ref. \cite{zhang_digital_2021} to retrieve the reconstructed image.  However, as also mentioned earlier, the fluctuations in our systems arise from variation in the transmission of the signal through the fog, not as fluctuations in background light that are not associated with the signal.

\section{Mitigation of fog fluctuations}
\begin{figure}[h!]
\centering\includegraphics[scale=1]{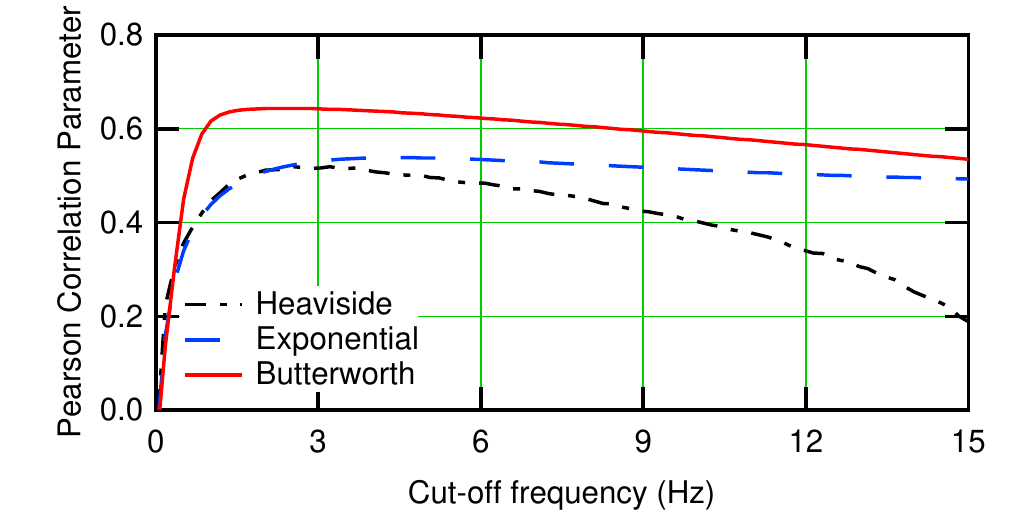}
\caption[]{Pearson correlation parameter for three different filter types: Heaviside, Exponential and Butterworth, as a function of cut-off frequency.}
\label{fig:pearson}
\end{figure}

As mentioned previously, to eliminate fog induced fluctuations, we decided to use a high-pass filter on the bucket signals. The computational nature of both CGI and CSI allows us to attempt different filters on the same data set, since the bucket values for imaging with and without fog are stored in the computer together with the corresponding speckle patterns. We can compare the high-pass filtered bucket values with fog to those measured without fog to determine how well this technique works. Of course this approach cannot yield a reconstruction completely identical to the one without fog since some functional lower frequencies are also removed. To compare the effect of different filters we employed a Pearson correlation \cite{pearson1896vii,taylor1997introduction} parameter between curves $x$ and $y$ given by

\begin{equation}
    r(x,y) = \frac{\sum(x-\bar{x})(y-\bar{y})}{\sqrt{\sum (x-\bar{x})^2 \sum (y-\bar{y})^2}}
\label{eqn:pearson}
\end{equation}
where $\bar{x}$ is the mean of $x$ and $\bar{y}$ is the mean of y. Here, $r(x,y) = 1$ for perfect correlation. Figure \ref{fig:pearson} shows the Pearson correlation parameter for Heaviside, exponential and forward-backward Butterworth filters in frequency space as a function of the cut-off frequency.

The exponential filter, given by $filt(f,f_{cut-off},n) = 1 - \exp \left ( -\left ( \frac{f}{f_{cut-off}}\right )^n \right )$ has its best result for $n=2$ and is only slightly better than the Heaviside filter at its peak. We obtained the best overall results, about 20\% improvement compared to other filters, with the Butterworth filter. The cut-off frequency for best correlation is only a couple of Hertz, which indicates that most of the useful bucket values will be untouched by the filter, resulting in potential reconstruction of the image in fog.

\begin{figure}[h!]
\centering\includegraphics[scale=0.6]{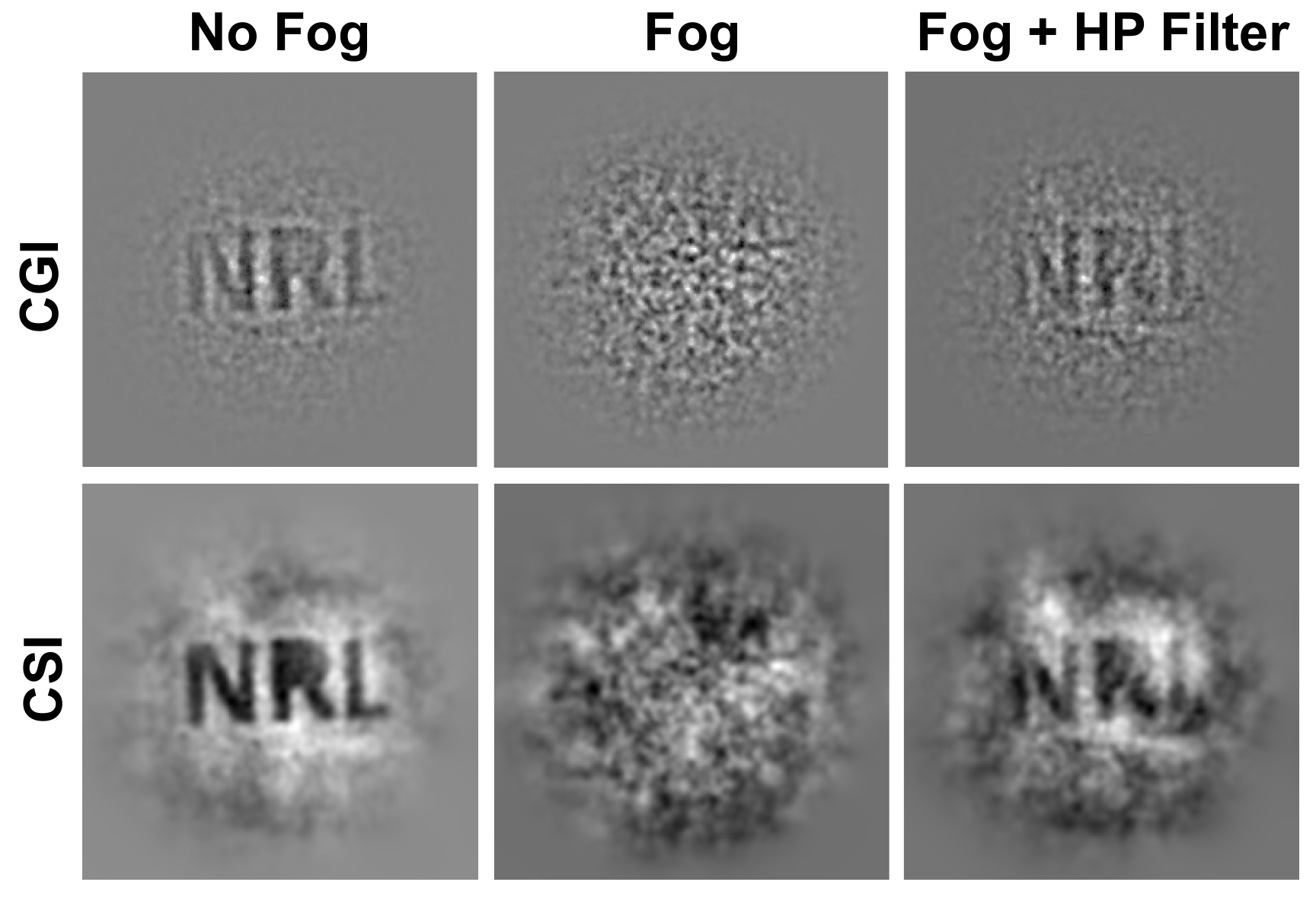}
\caption[]{CGI and CSI reconstructed images of the object without fog, with fog/no filter, and high-pass filtered bucket values using the Butterworth filter and 4000 speckle patterns.}
\label{fig:fog_filt}
\end{figure}

Figure \ref{fig:fog_filt} compares CGI (top row) and CSI (bottom row) reconstruction of the object without fog, with fog, and using the Butterworth high-pass filtered bucket values with fog in the first, second, and third columns, respectively. Unsurprisingly, both reconstructions fail for the case with the fog bucket values without the high-pass filter. Even though fog renders reconstruction impossible without filtering, applying the high-pass filtering to the bucket values results in a successful reconstruction for both CGI and CSI, shown in the third column of Figure \ref{fig:fog_filt}. The contrast and SNR of the image reconstructed using CSI is significantly enhanced compared to the CGI method.

\section{Conclusion}
We have demonstrated that time gating and high-pass filtering of the bucket values enable single-pixel structured image reconstruction in the laboratory fog chamber environment. We expect that the fluctuations due to fog in the field environment will occur on an even longer time scale. This should allow even better image reconstruction using our technique since fewer useful bucket values will be filtered out. Use of an orthogonal basis for projection, such as Hadamard patterns, can also help improve the efficacy and convergence of this technique for both CGI and CSI. 

\section*{Funding}
This work was supported by the Office of Naval Research.

\section*{Disclosures}
The authors declare that there are no conflicts of interest related to this paper.

\section*{Data availability}
Data underlying the results presented in this paper are not publicly available at this time.

\end{document}